\documentclass[conference]{IEEEtran}
\IEEEoverridecommandlockouts

\usepackage{amsthm}
\theoremstyle{plain}
\newtheorem{definition}{Definition}

\usepackage{caption}

\usepackage{booktabs, colortbl}       % Pretty tables

\usepackage{cite}           % For nicer citations (sorted and 1,2,3 -> 1-3)
\usepackage{comment}        % Introduces the comment environment to quickly (un)comment larger pieces of text
\usepackage{enumitem}

\usepackage{datetime}       % time access (mainly for compilation date+time in footer)
\yyyymmdddate

\usepackage{graphicx}
\graphicspath{{./figures/}}

\usepackage{hhline}         % For partial lines in tables
\usepackage{hyperref}       % For the handy \autoref{} command that prevends you from typing Section~\ref{}, also allows for global switching to/from Fig. and Figure
\usepackage{multirow}       % For table cells to span multipe rows

\usepackage{rotating}       % To rotate tables sideways on the page

\usepackage{subcaption}     % For subfigures (subfig(ure) packages are deprecated!)

\usepackage{amssymb, amsmath, eurosym}
\usepackage{tikz}           % TikZ is a handy package to 'draw' Petri nets and stuff inside the document and in a TeX style
\usepackage{xspace}     % For correct spacing in \ie \eg \etc macro's

\newcommand{\ie}{i.e.\@\xspace}

\newcommand{\eg}{e.g.\@\xspace}

\newcommand{\trace}[1]{\ensuremath{\langle #1 \rangle}} %Inline trace
\newcommand{\todo}[1]{\marginpar{\emph{\footnotesize#1}}} %For todo notes in the margin
\usepackage{color}

\newcommand{\diff}[1]{\textcolor{orange}{#1}}
\newcommand{\mcomm}[1]{\marginpar{\emph{\textcolor{orange}{\footnotesize#1}}}}
\newcommand{\remove}[1]{}
\newcommand{\csm}{CSM}
\newcommand{\csms}{CSMs}

\newcommand{\sm}{SM}

\newcommand{\sys}{\mathcal{M}}
\newcommand{\states}{S}
\newcommand{\state}{s}
\newcommand{\astates}{\overline{S}}
\newcommand{\trans}{T}
\newcommand{\inits}{b}
\newcommand{\fins}{f}
\newcommand{\pinits}[1]{b_{#1}}
\newcommand{\pfins}[1]{f_{#1}}
\newcommand{\pers}{\mathcal{A}}
\newcommand{\proj}{\Pi}
\newcommand{\sproj}{\pi_\proj}
\newcommand{\sproji}[1]{\pi_{\{#1\}}}
\newcommand{\forw}{\mathcal{F}}

\newcommand{\exec}{\sigma}
\newcommand{\execb}{\varsigma}
\newcommand{\execs}{\mathcal{L}}
\newcommand{\se}{e}
\newcommand{\statef}{\textsf{state}}
\newcommand{\pstatef}{\textsf{state}_i}
\newcommand{\timef}{\textsf{time}}
\newcommand{\freq}{\textsf{freq}}

\newcommand{\execduration}{\delta}
\newcommand{\sojourn}{\textsf{soj}}
\newcommand{\tfreq}{\textsf{freq}_\trans}
\newcommand{\ffreq}{\textsf{freq}_\filt}

\newcommand{\traces}{\textsf{Traces}}

\newcommand{\timed}{\mathbb{T}}%{\bbbt}

\newcommand{\items}{I}
\newcommand{\itemsets}{D}
\newcommand{\itemset}{d}
\newcommand{\execitemset}{\iota}
\newcommand{\exectransitemset}{\tau}

\newcommand{\package}{CSMMiner}
\newcommand{\plugin}{CSM Miner}

\newcommand{\supp}{\textsf{supp}}
\newcommand{\conf}{\textsf{conf}}
\newcommand{\lift}{\textsf{lift}}
\newcommand{\conv}{\textsf{convic}}
\newcommand{\cosi}{\textsf{cosine}}
\newcommand{\jacc}{\textsf{jaccard}}
\newcommand{\phim}{\phi}

\begin{document}

\title{Guided Interaction Exploration in Artifact-centric Process Models}

%\author{Maikel L. van Eck\thanks{This research was performed in the context of the IMPULS collaboration project of Eindhoven University of Technology and Philips: ``Mine your own body''.} \and Natalia Sidorova \and Wil M.P. van der Aalst}
%\institute{
%    Eindhoven University of Technology, The Netherlands\\
%        \email{\{m.l.v.eck,n.sidorova,w.m.p.v.d.aalst\}@tue.nl}
%}
\author{
\IEEEauthorblockN{Maikel L. van Eck\IEEEauthorrefmark{1}\thanks{\IEEEauthorrefmark{1}This research was performed in the IMPULS collaboration project of Eindhoven University of Technology and Philips: ``Mine your own body''.}, Natalia Sidorova, Wil M.P. van der Aalst}
\IEEEauthorblockA{Eindhoven University of Technology, The Netherlands\\
        Email: \{m.l.v.eck,n.sidorova,w.m.p.v.d.aalst\}@tue.nl}
}

\maketitle

\setcounter{footnote}{0}
\begin{abstract}
%Show compile date and time in margin for easier version management
%\todo{\today\ \currenttime}
Artifact-centric process models aim to describe complex processes as a collection of interacting artifacts.
Recent development in process mining allow for the discovery of such models.
However, the focus is often on the representation of the individual artifacts rather than their interactions.
Based on event data we can automatically discover composite state machines representing artifact-centric processes.
Moreover, we provide ways of visualizing and quantifying interactions among different artifacts.
For example, we are able to highlight strongly correlated behaviours in different artifacts.
The approach has been fully implemented as a ProM plug-in; the CSM Miner provides an interactive artifact-centric process discovery tool focussing on interactions.
The approach has been evaluated using real life data sets, including the personal loan and overdraft process of a Dutch financial institution.
\end{abstract}

\section{Introduction}\label{sec:introduction}

%\change{This is the introduction: page 1-2.}

%Paper motivation:

\begin{comment}
  \item Process discovery techniques can automatically produce models, which can be studied to identify unexpected or interesting process flows that differ from the process behaviour expected by a process expert or analyst.
  \item This is not trivial because process models can be complex, there are often multiple views on the same process, and outlier detection is difficult without a good definition of unexpected and interesting.
  \item It is even more difficult to identify interesting relations between process elements from two different models that belong to artifacts interacting in a bigger system.
  \item There exist measures in the fields of association rule mining and sequential pattern mining that attempt to quantify the interestingness of unexpected connections within data.
  \item These measures are based on probabilities, but not on semantics of states or activities
  \item This paper explores the value of these measures in the context of process artifact interaction and investigates whether results from related fields also hold in this context, i.e. how do the views of process (analysis) experts correlate with the measures of interestingness and can we learn e.g. a regression model to approximate an expert view to guide the exploration of mined models.
  \item With this information it is possible to create (tool) guidance for process analysts that highlights interesting relations in systems of interacting process artifacts.
\end{comment}

Process discovery is the automated creation of process models that explain the behaviour captured in event data~\cite{Aalst2016Book}.
These process models can be studied \eg to identify interesting process flows that differ from the process behaviour expected by a process expert or analyst.
However, complex process behaviour can result in unstructured process models, which makes them difficult and time-consuming to analyse.
Furthermore, there are often multiple views on the same process, and analysts do not always know what they are looking for.

One of the sources of complexity of discovered process models is that many process discovery approaches produce models that provide a monolithic view on the real process~\cite{Aalst2016Book,DBLP:conf/bpm/EckSA16}.
These models generally explain the behaviour of a process in terms of the life-cycle of a single process instance.
However, in reality a process instance may involve several interacting process objects or artifacts, each with their own life-cycle~\cite{DBLP:journals/ijcis/AalstBEW01,DBLP:journals/ijcis/PopovaFD15}.
For example, a procurement process with order and invoice objects, the behavioural process of a smart product with sensors that detect the product's state for different physical aspects, or the status of a single resource in terms of its status in the different processes it is involved in. %involved in different processes. %each modelled as artifacts.
%human being for whom the homeostatic processes of sleep and nutrition can be modelled as interacting artifacts.

Recently, it has become possible to automatically discover models for process artifacts and their behavioural interactions~\cite{DBLP:journals/tsc/LuNWF15,DBLP:conf/bpm/EckSA16,DBLP:journals/ijcis/PopovaFD15}.
These techniques produce individual process models for each artifact or perspective similar to traditional process discovery approaches.
The addition of artifact interaction enriches the individual models, connecting process elements from different artifact models.
Such information highlights \eg whether a specific state in one artifact coincides with the state of another artifact.

Artifact-centric techniques can provide more structured process models than traditional discovery approaches~\cite{DBLP:conf/bpm/EckSA16}.
However, decomposing the behaviour of a process into interacting artifacts does not necessarily make the overall process easier to understand.
Therefore, we present an approach to support the \emph{analysis of behavioural interactions between process artifacts}.
The goal is to find the most interesting or relevant interactions so that an analyst can inspect these first.
This helps process analysts faced with complex processes featuring artifacts interacting in a bigger system.

There are different ways to interpret the interaction of artifacts~\cite{DBLP:journals/tsc/LuNWF15,DBLP:conf/bpm/EckSA16,DBLP:conf/bpm/PopovaD13}.
We are interested in finding implications that given the occurrence of an element of one artifact-lifecycle provide information on the possible behaviour of other artifacts.
Process data generally does not explicitly contain these interactions or causal relations between artifact behaviour, so instead, we use information on \emph{correlations between artifact behaviour to obtain such insights}.

The analysis guidance involves the use of measures of interestingness to quantify artifact interactions.
Such measures have been developed in the field of association rule learning to quantify the relevance of relations between sets of items~\cite{DBLP:journals/is/TanKS04,DBLP:journals/expert/LiuHCM00}.
In this work we show how these measures can be defined in the context of process artifact interaction.
Based on these measures a ranking of artifact interactions can be presented to process analysts when inspecting process discovery results.
We have extended our artifact-centric process discovery tool, the \plugin~\cite{DBLP:conf/bpm/EckSA16a} in the ProM process mining framework, to support the explanation and analysis of interactions.

To evaluate the use of analysis guidance in practice we have used the developed tool with real life process data.
We discuss the results of this analysis and compare it to insights obtained by other researchers using traditional process mining approaches on the same data.
This evaluation shows that the analysis guidance provides insights into the overall process behaviour by highlighting interesting artifact interactions.

\begin{comment}
To evaluate the use of analysis guidance in practice we have performed a survey study.
In this study we investigated how the views of process analysts relate to the measures of interestingness for artifact interaction.
Participants were asked to rate process artifact interaction relations on usefulness and unexpectedness.
The goal of the study was to see if the subjective interests of process analysts can be approximated with objective measures of interestingness in order to provide an overview of the most relevant artifact interactions.

%Paper content:

\begin{comment}
    \item Introduce process discovery, systems of artifacts and their interaction
    \item Explain how the interaction between process artifacts relates to e.g. association rules and sequential patterns
    \item Discuss selection of measures (based on related work)
    \item Define measures of interestingness in process artifact system context
    \item Introduce evaluation method with a survey of process analysis experts that have been asked to rate process artifact interaction relations on usefulness and unexpectedness
    \item Results can be used to create a system to provide guidance when exploring systems of interacting process artifacts
\end{comment}

The remainder of this paper is structured as follows.
First, in \autoref{sec:related} we discuss related work on artifact-centric process mining and measures of interestingness.
In \autoref{sec:artifacts} we introduce a way to model processes representing artifact systems and define artifact interactions.
Then in \autoref{sec:metrics} we define measures of interestingness in the context of process artifacts.
We present the implementation of the analysis guidance in the \plugin~in \autoref{sec:implementation}.
We evaluate the tool using real life process data in \autoref{sec:evaluation}.
Finally, in \autoref{sec:conclusion} we present future work and conclusions. 

\section{Related Work}\label{sec:related}

%\change{This is the related work: page 4.}

A plethora of algorithms and tools for automated process discovery emerged over the last decade~\cite{Aalst2016Book}.
These produce models in various process model notations.
Several approaches have also been developed to take an object-oriented or artifact-centric view of process mining\cite{DBLP:journals/ijcis/AalstBEW01,DBLP:journals/ijcis/PopovaFD15}.
However, the number of techniques that can automatically discover the interactions between artifact models is limited~\cite{DBLP:journals/tsc/LuNWF15,DBLP:conf/bpm/EckSA16}.

There are different types of behavioural interaction between artifacts that can be mined from process execution data.
Like in monolithic process discovery, it is possible to establish causal dependencies between events that occur in different artifacts~\cite{DBLP:journals/tsc/LuNWF15}.
It is also possible to link a stage in one artifact lifecycle to stages in related artifact lifecycles by discovering synchronization conditions\cite{DBLP:conf/bpm/PopovaD13}.
Similarly, one can identify artifact interaction defined as the co-occurrence of states and transitions from different artifacts as part of the states and transitions of the entire process~\cite{DBLP:conf/bpm/EckSA16}.

The goal of the analysis of process artifacts and their interaction is to help the user understand complex behaviour by providing additional structure to the process through decomposition.
There are several other existing approaches in process mining to deal with model complexity.
Most process discovery tools have filtering options or sliders to adjust which activities and dependencies between activities are shown, often based on frequency information~\cite{Aalst2016Book}.
For some types of processes it is also possible to discover hierarchical process models that allow the analysis of a process at different levels of detail~\cite{DBLP:conf/caise/BoseVA11a}.
Trace clustering is a technique to decompose the process data of flexible processes with many different process instance variants that share little overlap in behaviour~\cite{DBLP:journals/tkde/WeerdtBVB13}.
The clustered process instances are used to mine a more limited model with fewer and stronger dependencies between activities.
However, all these approaches simplify the real behaviour shown by the data and hide information instead of using the complete information to guide the analyst.

Understanding the relations between artifacts and their effect on the overall process behaviour is a challenge~\cite{DBLP:journals/tsc/LuNWF15}.
For complex processes this requires the analysis of large numbers of possible artifact interactions, many of which are not interesting.
This problem is related to the problem in association rule learning that association rule mining algorithms produce large numbers of rules that are not equally relevant~\cite{DBLP:conf/edm/BazalduaBP14,DBLP:journals/expert/LiuHCM00,DBLP:journals/is/TanKS04}.
A solution in association rule learning for this problem involves the quantification of the interestingness of the association rules using specific measures of interestingness.

\section{Modelling of Artifact Systems}\label{sec:artifacts}

In this work we use the notion of state machines to model processes representing artifact systems and the life-cycles of artifacts as presented in~\cite{DBLP:conf/bpm/EckSA16}.
We developed the \plugin~to support such models~\cite{DBLP:conf/bpm/EckSA16a}.

Regarding notation, we write $\exec_k$ for the $k$-th element of a sequence $\exec \in \states^*$ of elements from some set $\states$, and $|\exec|$ denotes the length of $\exec$.
We write ${\state \in \exec}$ if $\state = \exec_k$ for some $k$ and $\exec\trace{\state,\dots,\state'}$ for the concatenation of $\exec$ with sequence $\trace{\state,\dots,\state'}$.
Additionally, for $\state \in \states_1 \times \dots \times \states_n$ we write $\state(i)$ for the value of the $i$-th component of $s$ ($i \in \{1, \dots, n\}$).

\subsection{Composite State Machines}\label{sec:statemachines}

\begin{comment}
We define a \emph{State Machine} (\sm) as follows:

\begin{definition}\label{def:statemachine}
A \emph{State Machine} $\sys$ is a tuple $(\states, \trans, \inits, \fins)$ where
$\states$ is a set of states,
$\trans \subseteq (\states \cup \{ \inits \}) \times (\states \cup \{ \fins \})$ is the set of transitions,
$\inits \notin \states$ is the initial source state,
and $\fins \notin \states$ is the final sink state.
%\diff{$(\state,\state') \in \trans$ is also denoted as $(\state\rightarrow\state')$.}
We define $\astates= \states \cup \{ \inits, \fins \}$.
\end{definition}
\end{comment}

A process consisting of a number of interacting artifacts is called an artifact system, and we model its behaviour as a \emph{Composite State Machine} (\csm).
The state of a \csm~is defined as the composition of the states of its artifacts, \ie it is a vector of states.
The set of all possible states of a \csm~is a subset of the cartesian product of the sets of states of its artifacts, as not all combinations of artifact states are necessarily possible.
Each transition in a \csm~represents a change in the state of at least one artifact; we do not allow self loops.
Formally:

\begin{definition}\label{def:system}
A \emph{Composite State Machine} $\sys = (\states, \trans, \inits, \fins)$ is a model of a process with $n$ artifacts where
$\states \subseteq (\states_1 \times \dots \times \states_n)$ is a set of states, with $S_1, \ldots, S_n$ the sets of artifact states,
$\inits = (\pinits{1} , \dots , \pinits{n})$ is the initial source state,
$\fins =\nobreak (\pfins{1} , \dots , \pfins{n})$ is the final sink state,
$\trans \subseteq (\states \cup \{ \inits \}) \times (\states \cup \{ \fins \})$ is the set of transitions,
and $\forall (\state, \state') \in \trans : \state \neq \state'$.
We define $\astates= \states \cup \{ \inits, \fins \}$ and $\astates_i= \states_i \cup \{ \pinits{i}, \pfins{i} \}$ for $i \in \{ 1, \dots, n \}$.
\end{definition}

The explicit initial and final states have no incoming and outgoing transitions, respectively.
They are not true states: they only mark the points in time where a process instance begins and finishes.
As a special case, we call a \csm~with only one artifact an \emph{Artifact Model}, which represents the behaviour of the artifact in isolation.

We can project a \csm~onto a specific subset of its artifacts to focus only on their behaviour.
A \emph{\csm~Projection} is obtained by reducing the cartesian product of each state to the given subset of artifacts, merging the identical states, and omitting unnecessary transitions and self loops.
%Since transitions represent state changes, two states $s_i,s'_i$ of an Artifact Model $\pers_i$ are connected by a transition iff there is a transition from some state $s$ to a state $s'$ in the \csm~that changes the value of the i-th state component from $s_i$ to $s'_i$.
As transitions represent state changes, two states of a projection are only connected by a transition if there is a transition in the \csm~whose source and target are reduced to these different states.
%As a special case, a \csm~can be projected onto a single \emph{Artifact Model}, which represents the behaviour of a single artifact in isolation.

\begin{figure*}[btp]
\centering
\begin{subfigure}[b]{0.6\textwidth}
	\centering
    \includegraphics[width=0.95\textwidth, trim=0 0 0 0]{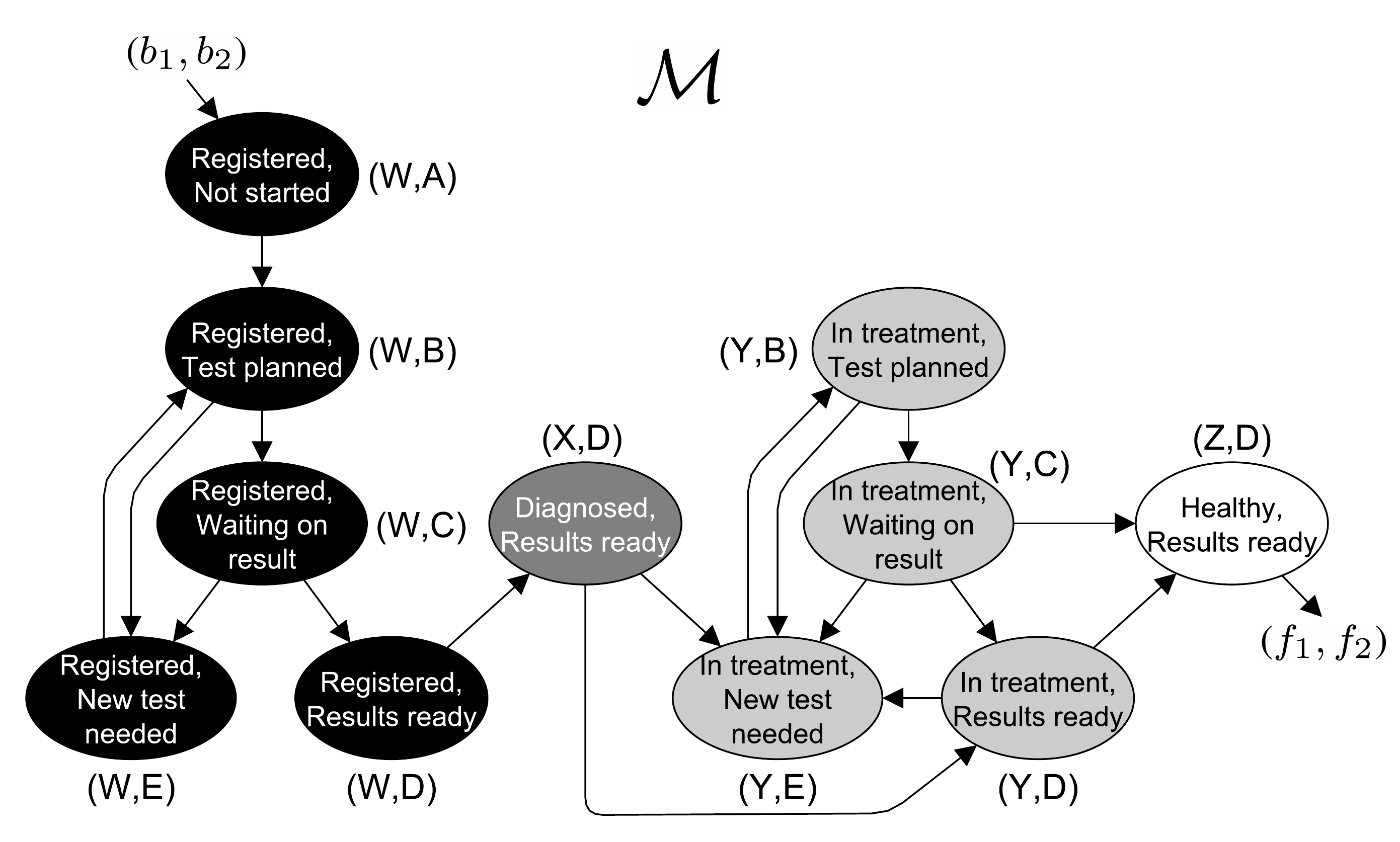}
	%\caption{}
	\label{fig:compositeModel}
\end{subfigure}
\begin{subfigure}[b]{0.13\textwidth}
	\centering
	\includegraphics[width=0.99\textwidth, trim=10 55 0 0]{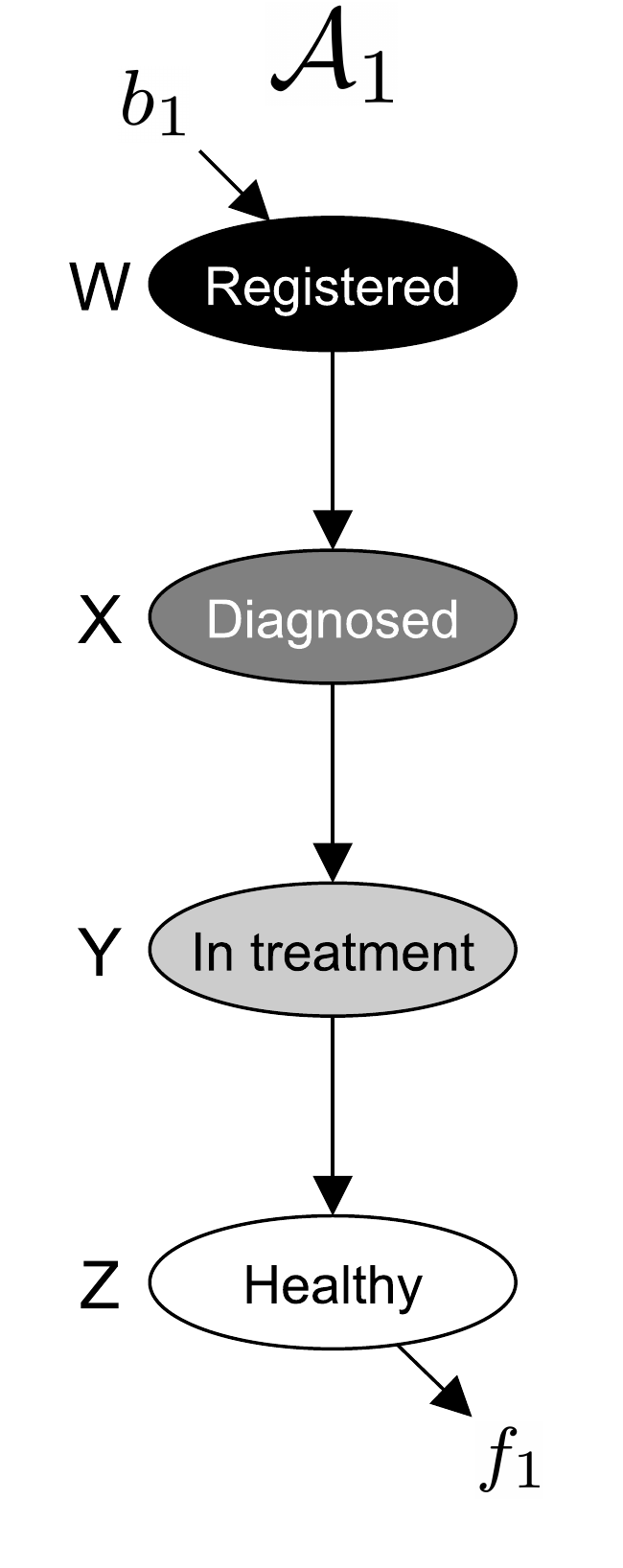}
	%\caption{}
	\label{fig:treatmentModel}
\end{subfigure}
\begin{subfigure}[b]{0.23\textwidth}
	\centering
	\includegraphics[width=0.95\textwidth, trim=0 60 10 0]{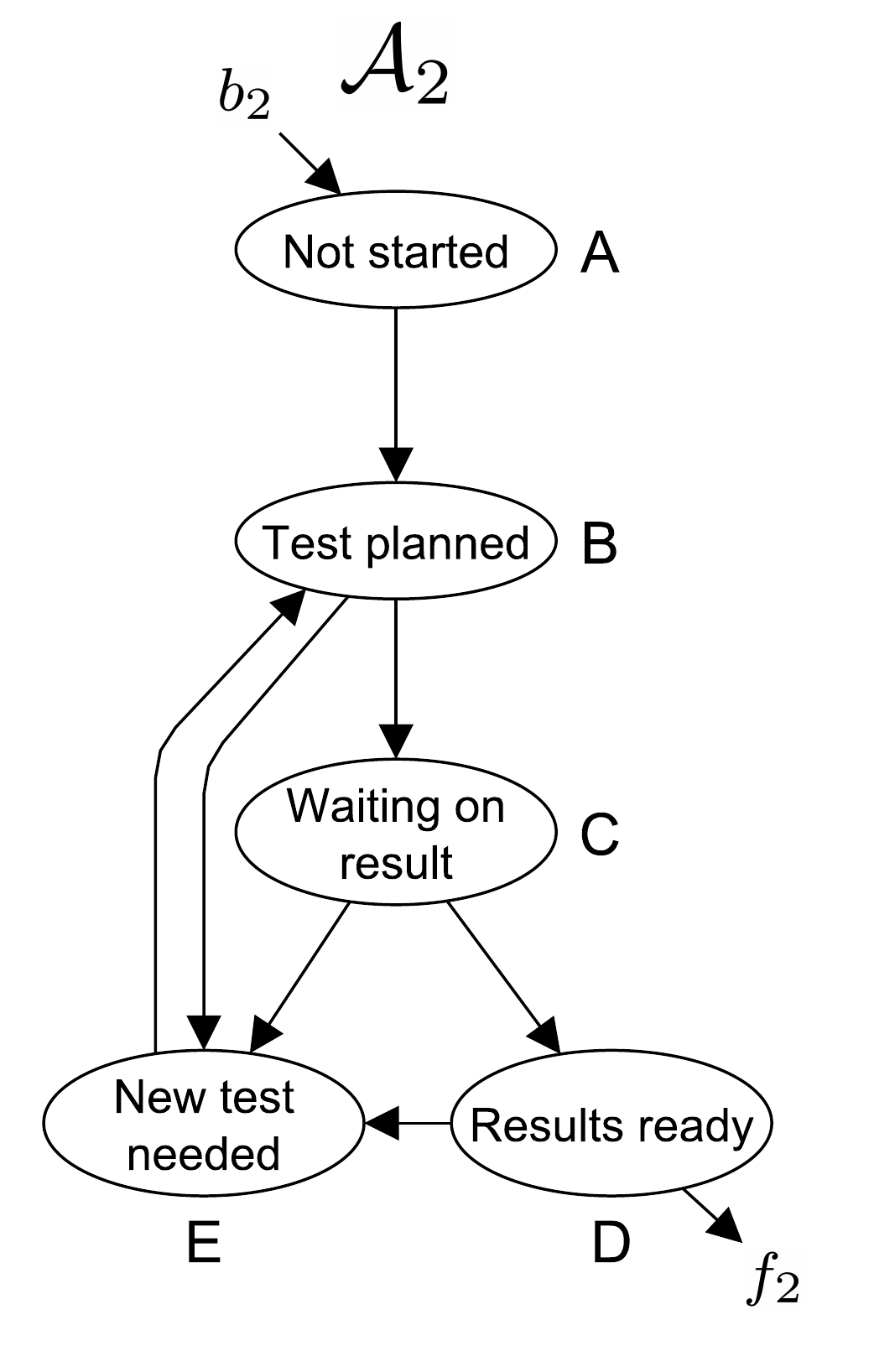}
	%\caption{}
	\label{fig:testModel}
\end{subfigure}

\caption{A model $\sys$ of a simple healthcare process and its two artifact models $\pers_1$ and $\pers_2$. Every state in the process is a combination of a state from each artifact.}
\label{fig:runningExample}
\end{figure*}

\begin{definition}\label{def:systemprojection}
Given a \csm~$\sys$ %= (\states, \trans, \inits, \fins)$, with $\states \subseteq (\states_1 \times \dots \times \states_n)$,
and an ordered subset of indices $\proj = \{ i_1, \dots, i_m \} \subseteq \{ 1, \dots, n \}$, with $i_1 < \nobreak i_2 < \dots < i_m$, %$\forall j \in\nobreak \{ 1, \dots, m-\nobreak1 \} : i_j < i_{j+1}$,
we define the state projection function $\sproj:~(\astates_1 \times\nobreak \dots \times\nobreak \astates_n) \rightarrow (\astates_{i_1} \times \dots \times \astates_{i_m})$
as follows: $\forall \state \in \astates, i_j \in \proj : (\sproj(\state))(j) = \state(i_j)$.
A \emph{\csm~Projection} of $\sys$ on $\proj$, $\sys^\proj = (\states^\proj, \trans^\proj, \inits^\proj, \fins^\proj)$, is defined as:

\vspace{5pt}
\begin{tabular}{>{$}l<{$} @{${}={}$} >{$}l<{$}}
    \states^\proj & \{ \sproj(\state) | \state \in \states \}, \\
    \trans^\proj & \{ (\sproj(\state), \sproj(\state')) | (\state, \state') \in \trans \wedge \sproj(\state) \neq \sproj(\state') \}, \\
    \inits^\proj & \sproj(\inits), \\
    \fins^\proj & \sproj(\fins).
\end{tabular}

\vspace{5pt}\noindent
%We define \emph{Artifact Model} $\pers_i$ as the projection $\sys^{\{i\}}$ of $\sys$ on $\{i\}$. %, \ie a \csm~of the behaviour of a single artifact $(n = 1)$.
The \emph{Artifact Model} $\pers_i$ is defined as the projection $\sys^{\{i\}}$ of $\sys$ on $\{i\}$.
\end{definition}

Note that the projection of a \csm~is itself again a \csm, modelling only the behaviour of the artifacts projected on.

\begin{comment}\label{def:artifact}
\emph{Artifact Model} $\pers_i$ of a \csm~$\sys = (\states, \trans, \inits, \fins)$, with $\states \subseteq (\states_1 \times \dots \times \states_n)$ and $i \in \{1, \dots, n\}$,
is a state machine $\pers_i = (\states_i,\trans_i, \pinits, \pfins)$ such that:

\vspace{5pt}
\begin{tabular}{>{$}l<{$} @{${}={}$} >{$}l<{$}}
    \trans_i & \{ (\state(i), \state'(i)) | (\state, \state') \in \trans \wedge \state(i) \neq \state'(i) \} \\
    \pinits & \{s(i)|s\in \inits\} \\
    \pfins & \{s(i)|s\in \fins\}
\end{tabular}
\end{comment}

In \autoref{fig:runningExample} we present a simple healthcare process, which we use as a running example.
This process (model $\sys$) has two distinct perspectives or artifacts:
the status of the patient being treated (model $\pers_1$), and the status of lab tests of the patient (model $\pers_2$).
%\change{The initial states are marked with an incoming arrow and the final states are marked with an outgoing arrow.}
The artificial initial and finial states are marked without border.

%\autoref{fig:runningExample} shows an example process of X composed of two interacting artifacts: Y and Z.
%Explain behaviour of individual artifacts and composition.

The healthcare process starts when the patient is registered, after which a lab test is planned to diagnose the patient.
If the patient misses their appointment or if the results are inconclusive, then a new test is planned, but if the test results are ready then the treatment can proceed.
During the treatment additional tests may be required, until the patient is healthy again and the process ends.
Note that the composite process is smaller than the cartesian product of the artifacts ($4 \times 5 = 20$ states) because not all state combinations can be observed due to interdependencies.
For example, once the patient is healthy no extra lab tests are needed.
Such dependencies between artifacts can be interesting to analyse.

\subsection{Process Execution Data}\label{sec:executiondata}

The \csm~models as introduced above provide only limited insights into the dependencies and interaction between the artifacts whose behaviour makes up the process of the artifact system.
There are no expected sojourn times for the different states or frequencies for transitions.
For the process in \autoref{fig:runningExample} an analyst could be interested \eg in the average time spent \emph{Waiting on result~(C)} while the patient is \emph{In treatment~(Y)} or the difference in probability of transitioning to \emph{New test needed~(E)} before and after the patient is \emph{Diagnosed~(X)}.
To enrich the model with such information, we require a collection of process execution data.

In this work we assume the availability of both a \csm~of the process of interest and a matching collection of process instance data consisting of execution sequences of the process.
Each \emph{State Entry} in an \emph{Execution Sequence}, or trace, specifies the new state of the artifact system at a certain point in time.
A collection of execution sequences together form a \emph{Log}.
Given a log, a \csm~can be discovered that matches the execution sequences in the log~\cite{DBLP:conf/bpm/EckSA16}.

\begin{definition}\label{def:execution}
Let $\sys$ be a \csm~and $\timed$ a time domain.
We call $\se \in (\astates \times \timed)$ a \emph{State Entry}.
Function $\statef(\se)$ returns the state, $\timef(\se)$ returns the time, and $\pstatef(\se) = \sproji{i}(\statef(\se))$~returns the state projection of the state entry $e$.

%Let $\sys = (\states, \trans, \inits, \fins)$ be a \csm, with $\states \subseteq (\states_1 \times \dots \times \states_n)$ and $\pers_1, \dots, \pers_n$ its artifacts.
%Let $\exec_k \in (\states \times \timed)$ be a \emph{State Entry} with a timestamp from time domain $\timed$.
%Function $\statef(\exec_k)$ returns the state of $\sys$ at time $\timef(\exec_k)$, and $\pstatef(\exec_k)$~returns the state of artifact $\pers_i$.

$\exec \in (\astates \times \timed)^*$ is an \emph{Execution Sequence} of $\sys$ iff:
\begin{itemize}
    \item $\statef(\exec_1) = \inits$,
    \item $\statef(\exec_{|\exec|}) = \fins$,
    \item $(\statef(\exec_k), \statef(\exec_{k+1})) \in \trans$ for $k \in \{1, \dots, |\exec|-1\}$,
    \item $\timef(\exec_1) = \timef(\exec_{2})$, and
    \item $\timef(\exec_k) < \timef(\exec_{k+1})$ for $k \in \{2, \dots, |\exec|-1\}$.
\end{itemize}
The set $\traces_\sys$ is the set of all possible execution sequences of $\sys$.
A \emph{Log} $\execs_\sys : \traces_\sys \rightarrow \mathbb{N}$ is a multiset of execution sequences.
%Function $\freq_\sys(\exec)$ returns the frequency of $\exec \in \traces_\sys$.
\end{definition}

\begin{figure*}[bt]
\centering
\includegraphics[width=0.99\textwidth]{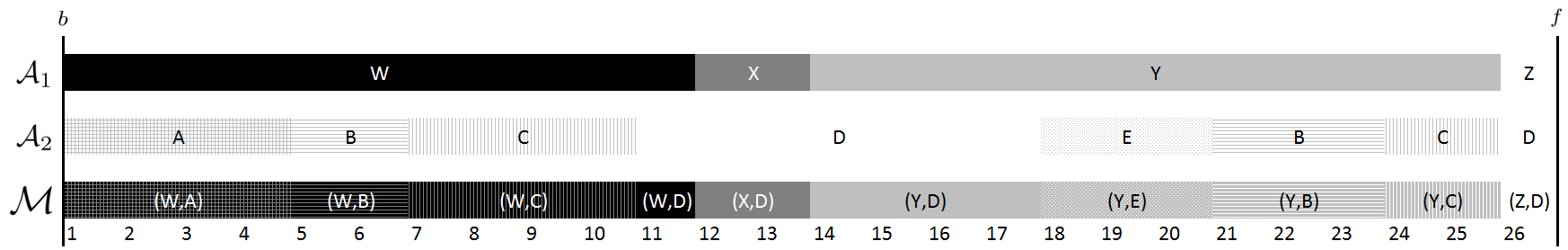}
\caption{An execution sequence for the running example process from \autoref{fig:runningExample}.}
\label{fig:log}
\end{figure*}

An example of an execution sequence for the \csms~from \autoref{fig:runningExample} is provided in \autoref{fig:log}.
Note that no time is spent in the artificial initial state $\inits$, representing the beginning of the known execution, but it is included in execution sequences to enable the calculation of the frequency of the different possible ways to start a process.
Artificial final state $\fins$ represents the point in time after which the process instance finished and the state is unknown.

We can use the time information in an execution sequence to calculate the time spent in a given state.
By aggregating the durations of state entries over a log the models can be enriched with sojourn time statistics for each state.
Similar to state sojourn times, we can also count the number of transitions occurring in a log.
These numbers can be used to annotate the transitions in the process models with frequency statistics.

\begin{table*}
\centering
\caption{The state entries of the execution sequence $\exec$ of $\sys$ from \autoref{fig:log}, the sequence projected on the first artifact $\exec' = \sproji{1}(\exec)$, and the sequence projected on the second artifact $\exec'' = \sproji{2}(\exec)$.}
\label{tab:executionSequence}

\begin{tabular}
	{ c | l | c }
	$k$ & \multicolumn{1}{c|}{$\exec_k$} & $\execduration(\exec_k)$ \\
    \hline
    1 & (($\pinits{1},\pinits{2}$), 1-1-'17) & 0 \\
    2 & ((W,A), 1-1-'17) & 4 \\
    3 & ((W,B), 5-1-'17) & 2 \\
    4 & ((W,C), 7-1-'17) & 4 \\
    5 & ((W,D), 11-1-'17) & 1 \\
    6 & ((X,D), 12-1-'17) & 2 \\
    7 & ((Y,D), 14-1-'17) & 4 \\
    8 & ((Y,E), 18-1-'17) & 3 \\
    9 & ((Y,B), 21-1-'17) & 3 \\
    10 & ((Y,C), 24-1-'17) & 2 \\
    11 & ((Z,D), 26-1-'17) & 1 \\
    12 & (($\pfins{1},\pfins{2}$), 27-1-'17) & 0 \\
\end{tabular}
~
\begin{tabular}
    { c | c | c }
    $l$ & $\exec'_l$ & $\execduration(\exec'_l)$ \\
    \hline
    1 & ($\pinits{1}$, 1-1-'17) & 0 \\
    2 & (W, 1-1-'17) & 11 \\
    3 & (X, 12-1-'17) & 2 \\
    4 & (Y, 14-1-'17) & 12 \\
    5 & (Z, 26-1-'17) & 1 \\
    6 & ($\pfins{1}$, 27-1-'17) & 0 \\ \multicolumn{1}{c}{} \\ \multicolumn{1}{c}{} \\ \multicolumn{1}{c}{} \\ \multicolumn{1}{c}{} \\ \multicolumn{1}{c}{} \\ \multicolumn{1}{c}{} \\

    %\multicolumn{1}{c}{} \\
\end{tabular}
~
\begin{tabular}
    { c | c | c }
    $m$ & $\exec''_m$ & $\execduration(\exec''_m)$ \\
    \hline
    1 & ($\pinits{2}$, 1-1-'17) & 0 \\
    2 & (A, 1-1-'17) & 4 \\
    3 & (B, 5-1-'17) & 2 \\
    4 & (C, 7-1-'17) & 4 \\
    5 & (D, 11-1-'17) & 7 \\
    6 & (E, 18-1-'17) & 3 \\
    7 & (B, 21-1-'17) & 3 \\
    8 & (C, 24-1-'17) & 2 \\
    9 & (D, 26-1-'17) & 1 \\
    10 & ($\pfins{2}$, 27-1-'17) & 0 \\ \multicolumn{1}{c}{} \\ \multicolumn{1}{c}{} \\
\end{tabular}

\end{table*}

\begin{definition}\label{def:stateDuration}
Let $\exec_k$ be a state entry of an execution sequence $\exec \in \traces_\sys$ of \csm~$\sys$.
The state entry's duration is given by:
\[
    \execduration(\exec_k) =
    \begin{cases}
        \text{\timef}(\exec_{k+1}) - \text{\timef}(\exec_k), & \text{if } 1 \leq k < |\exec| \\
        0, & \text{if } k = |\exec|
    \end{cases}
\]
The total sojourn time of a state $\state \in \states$ for a log $\execs_\sys$ is:
\[
    \sojourn(\state, \execs_\sys) = \sum_{\exec \in \execs_\sys} \Big( \sum_{\{ k | \statef(\exec_k) = \state \}} \execduration(\exec_k) \Big) * \execs_\sys(\exec)
\]
The frequency of a transition $(\state,\state') \in \trans$ for a log $\execs_\sys$ is:
\begin{align*}
    & \tfreq((\state,\state'),\execs_\sys) = \\
    & \quad \sum_{\exec \in \execs_\sys} \big|\{ k | \statef(\exec_k) = \state \wedge \statef(\exec_{k+1}) = \state' \}\big| * \execs_\sys(\exec)
\end{align*}
\end{definition}

%\autoref{tab:executionSequence} shows the state entries and their corresponding durations for an execution sequence of the example process.

An execution sequence of a \csm~can also be projected onto a subset of its artifacts such that it is an execution sequence of the matching projected \csm.
The projection abstracts from state entries where the state of the specified artifacts does not change from the previous state entry, as these entries no longer represent transitions in the projected process model.
With such projections we can calculate sojourn and frequency statistics to enrich projected \csms~as before.

\begin{definition}\label{def:executionprojection}
Let $\sys$ be a \csm, $\proj$ a set of artifact indices, and $\sproj$ a state projection function.
%An \emph{Execution Sequence Projection} $\exec^\proj$ is a projection of $\exec \in \traces_\sys$ such that $\exec^\proj \in \traces_{\sys^\proj}$ and $\forall k \in \{1, \dots, |\exec|\}$ :
%\begin{align*}
%    \exists & l \in \{1, \dots, |\exec^\proj|\} : \\
%    & \qquad \sproj(\statef(\exec_k)) = \statef(\exec^\proj_l) \wedge {} \\
%    & \qquad \Big( \timef(\exec_k) = \timef(\exec^\proj_l) \vee \exists k' \in \{1, \dots, k-1\} : \\
%    & \qquad\qquad \timef(\exec_{k'}) = \timef(\exec^\proj_l) \wedge {} \\
%    & \qquad\qquad \forall k'' \in \{k', \dots, k-1 \} : \\
%    & \qquad\qquad\qquad \sproj(\statef(\exec_k)) = \sproj(\statef(\exec_{k''}))\Big)
%\end{align*}
We lift the application of projection function $\sproj$ to sequences $\exec \in \traces_\sys$ so that $\sproj(\exec) \in \traces_{\sys^\proj}$.
We define $\sproj(\exec)$ recursively:

If $\exec = \trace{}$ then $\sproj(\exec) = \trace{}$, and if $\exec = \trace{\se}$, with $\se \in (\astates \times \timed)$, then $\sproj(\exec) = \trace{(\sproj(\statef(\se)),\timef(\se))}$.
For an execution sequence $\exec\trace{\se_{1},\se_2}$,
\begin{align*}
%& Basis \\
%& \sproj(\exec) =
%\begin{cases}
%    \trace{} & \text{if } |\exec| = 0 \\
%    \trace{(\sproj(\statef(\se)),\timef(\se))} & \text{if } \exec = \trace{\se}, \\
%    & \quad \se \in (\astates \times \timed) \\
%\end{cases} \\
%& \text{For an execution sequence}~\exec\trace{\se_{1},\se_2}, \\
& \sproj(\exec\trace{\se_{1},\se_2}) =
\begin{cases}
    \sproj(\exec\trace{\se_{1}}), & \text{if } \sproj(\statef(\se_{1})) = \\
    & \quad \sproj(\statef(\se_2)) \\
    \sproj(\exec\trace{\se_{1}})\sproj(\trace{\se_2}), & \text{otherwise} \\
\end{cases}
\end{align*}

%An \emph{Execution Sequence Projection} $\exec^\proj$ is a projection of $\exec \in \traces_\sys$ such that $\exec^\proj \in \traces_{\sys^\proj}$ \diff{(Recursive definition)}:
%\[
%\exec^\proj =
%\begin{cases}
%    \emptyset & \text{if } |\exec| = 0 \\
%    (\sproj(\statef(\se)),\timef(\se)) & \text{if } \exec = \se \\
%    (\exec'\se_{k-1})^\proj & \text{if } \exec = \exec'\se_{k-1}\se_k \wedge {} \\
%    & \quad \sproj(\statef(\se_{k-1})) = \sproj(\statef(\se_k)) \\
%    (\exec'\se_{k-1})^\proj(\sproj(\statef(\se_k)),\timef(\se_k)) & \text{if } \exec = \exec'\se_{k-1}\se_k \wedge {} \\
%    & \quad \sproj(\statef(\se_{k-1})) \neq \sproj(\statef(\se_k)) \\
%\end{cases}
%\]

A \emph{Log Projection} $\execs_\sys^\proj: \traces_{\sys^\proj} \rightarrow \mathbb{N}$ of a log $\execs_\sys$ is a multiset of execution sequences such that:
$\forall \execb \in \traces_{\sys^\proj} : \execs_\sys^\proj(\execb) = \sum_{\exec \in \execs_\sys : \execb = \sproj(\exec)} \execs_\sys(\exec)$.
\end{definition}

\autoref{tab:executionSequence} shows an execution sequence $\exec$ of the running example process and its projections $\sproj(\exec)$ for $\proj = \{ 1 \}$ and $\proj = \{ 2 \}$, together with their corresponding durations.

\begin{figure*}[btp]
\centering
\begin{subfigure}[b]{0.6\textwidth}
	\centering
    \includegraphics[width=0.95\textwidth, trim=0 0 0 0]{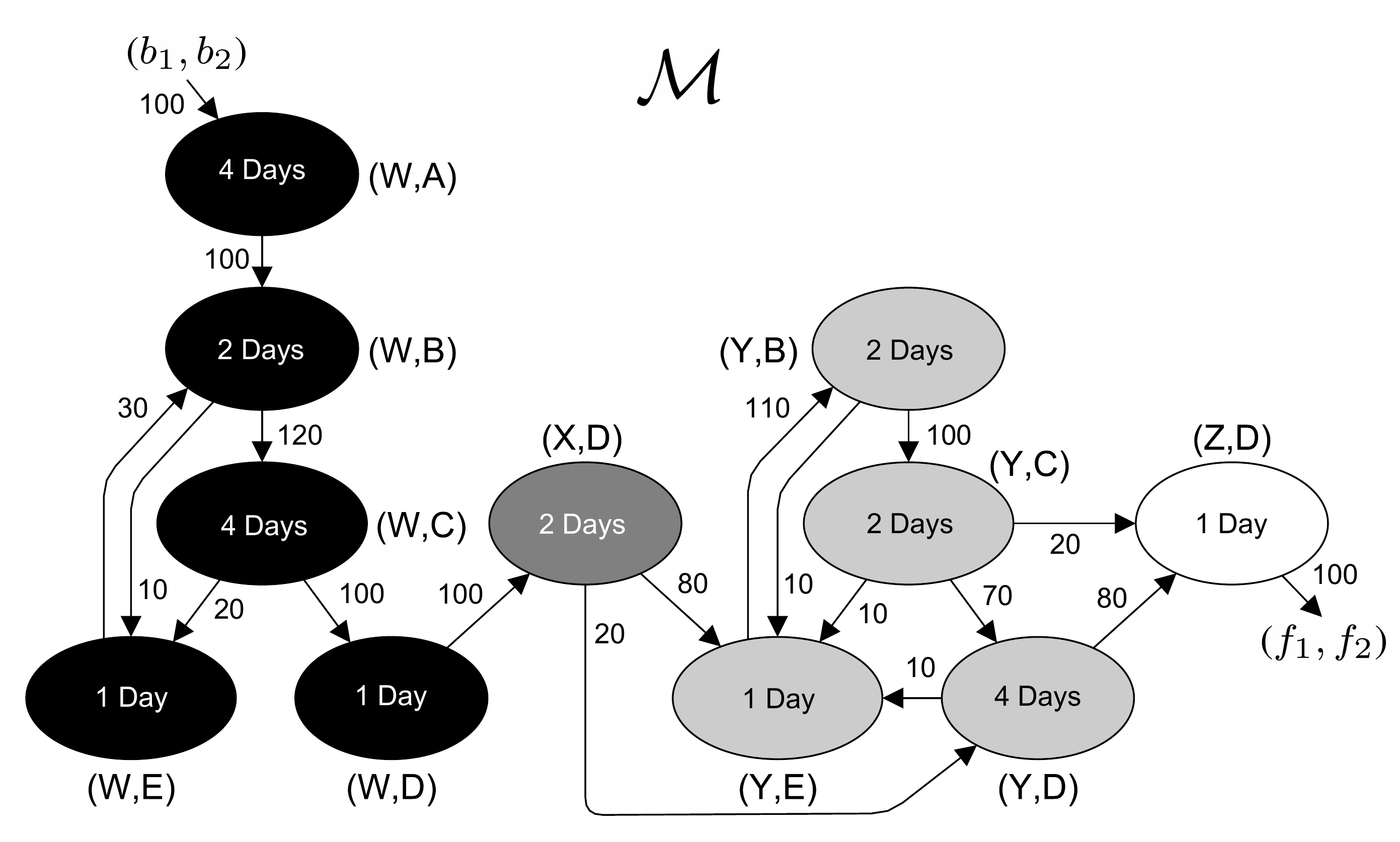}
	%\caption{}
	\label{fig:annotatedCompositeModel}
\end{subfigure}
\begin{subfigure}[b]{0.13\textwidth}
	\centering
	\includegraphics[width=0.99\textwidth, trim=10 55 0 0]{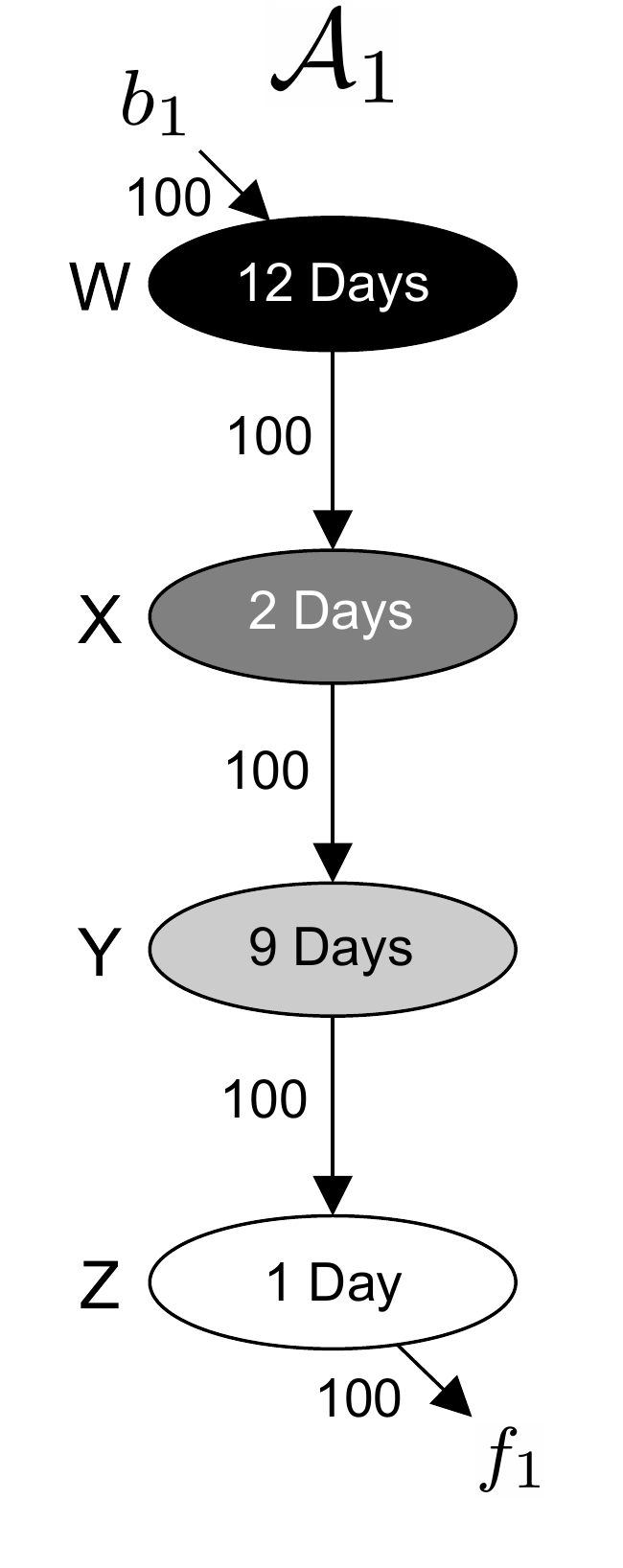}
	%\caption{}
	\label{fig:annotatedTreatmentModel}
\end{subfigure}
\begin{subfigure}[b]{0.23\textwidth}
	\centering
	\includegraphics[width=0.95\textwidth, trim=0 60 10 0]{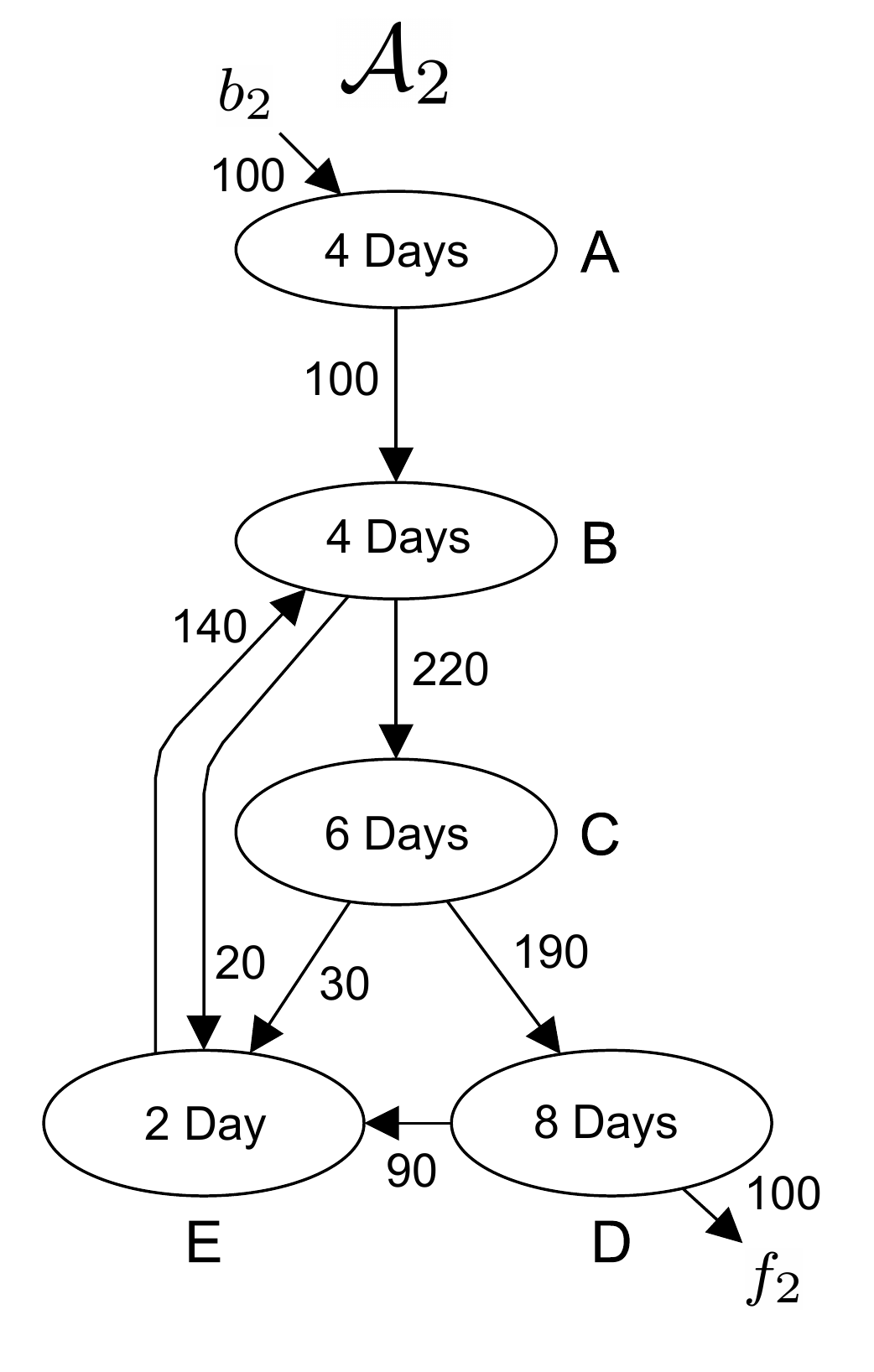}
	%\caption{}
	\label{fig:annotatedTestModel}
\end{subfigure}

\caption{The models of the healthcare process from \autoref{fig:runningExample} annotated with transition frequencies and average state sojourn times per trace.}
\label{fig:annotatedRunningExample}
\end{figure*}

The information in a collection of execution sequences can be used to enrich a \csm~and its projections with state sojourn statistics and transition frequencies as described above.
\autoref{fig:annotatedRunningExample} shows the running example process of \autoref{fig:runningExample} annotated with frequency and average sojourn time information.
Process execution data can also be used for the identification of relations between artifact model elements and the calculation of measures of interestingness for such relations.

\subsection{Artifact Interaction}\label{sec:interaction}

Given a \csm~$\sys$ with multiple artifacts and a log $\execs_\sys$, we want to find interesting artifact interactions that are a part of the artifact system behaviour.
For example, if the state of an artifact cannot be advanced until a certain state in a different artifact has been reached then this may represent a bottleneck in the overall process.
Similarly, the probability of making specific choices at a decision point in one artifact may be affected by the state of another artifact.
The executions in a log do not explicitly describe such causal dependencies between the behaviour of different artifacts, but we can infer \emph{correlations} between sets of artifact states or transitions.
Based on this, we distinguish three types of artifact interaction: \emph{state co-occurrence}, \emph{transition co-occurrence} and \emph{forward-looking co-occurrence}.

We focus here only on the interaction between pairs of artifacts, but the interaction definitions can be generalised to involve sets of artifacts.
%The \csm~models and a log of execution data for a pair of artifacts can be obtained by projecting the \csm~and the log onto the two artifacts of current interest, abstracting from other behaviour.
We formulate each interaction as an \emph{implication} $(X \Rightarrow Y)$ between two statements regarding the states or execution behaviour of the artifacts.

\emph{State co-occurrence} $(\state_i \Rightarrow_\states \state_j)$ is defined as the conditional probability that artifact model $\pers_j$ is in state $\state_j$ given that artifact model $\pers_i$ is in state $\state_i$.
From the execution sequences in a log we can determine the strength of this interaction in the observed data.
It is calculated as the amount of time the system state contains both states compared to the total time spent in $\state_i$.

\begin{definition}\label{def:stateinteraction}
Let $\sys$ be a \csm~with artifacts $i$ and $j$, $\state_i \in \states_i$~and $\state_j \in \states_j$.
The strength of the state co-occurrence $(\state_i \Rightarrow_\states \state_j)$ is defined as:
\[
\hat{P}_\states(\state_i \Rightarrow_\states \state_j) = \frac{\sojourn((\state_i,\state_j), \execs_\sys^{\{ i,j \}})}{\sojourn(\state_i, \execs_\sys^{\{ i \}})}
\]
\end{definition}

In \autoref{fig:annotatedRunningExample} we can see that the average time spent \emph{In treatment~(Y)} given that the lab is \emph{Waiting on result~(C)} is 2 days, while the average time spent \emph{Registered~(W)} given that the lab is \emph{Waiting on result~(C)} is 4 days.
So, the interaction $(C \Rightarrow_\states W)$ is a stronger co-occurrence ($\frac{4}{6}$) than the interaction $(C \Rightarrow_\states Y)$ ($\frac{2}{6}$).

\emph{Transition co-occurrence} $((\state_i,\state'_i) \Rightarrow_\trans (\state_j,\state'_j))$ is defined as the conditional probability that, given that $\pers_i$ is in a transition from $\state_i$ to $\state'_i$, $\pers_j$ has a state $\state_j$ before and a state $\state'_j$ after the transition.
If $\state_j = \state'_j$ this co-occurrence specifies the state of $\pers_j$ during the given transition in $\pers_i$, but if they differ then it specifies a transition in $\pers_j$ that co-occurs with the transition in $\pers_i$.
The strength of this interaction is calculated as the number of times we observe transitions for which both the condition and the consequence hold divided by the total number of observed transitions for which the condition holds.

\begin{definition}\label{def:transitioninteraction}
Let $\sys$ be a \csm~with artifacts $i$ and $j$, $\state_i, \state'_i \in\nobreak \astates_i$, $\state_i \neq \state'_i$, and $\state_j, \state'_j \in \astates_j$.
The strength of the transition co-occurrence $((\state_i,\state'_i) \Rightarrow_\trans (\state_j,\state'_j))$ is defined as:
\[
\hat{P}_\trans((\state_i,\state'_i) \Rightarrow_\trans (\state_j,\state'_j)) = \frac{\tfreq(((\state_i,\state_j),(\state'_i,\state'_j)),\execs_\sys^{\{ i,j \}})}{\tfreq((\state_i,\state'_i),\execs_\sys^{\{ i \}})}
\]
\end{definition}

In \autoref{fig:annotatedRunningExample} there are three types of transitions from \emph{Waiting on result~(C)} to \emph{Result ready~(D)}: while the patient is \emph{Registered~(W)} (100 times), while the patient is \emph{In Treatment~(Y)} (70 times), and simultaneously together with a transition from \emph{In treatment~(Y)} to \emph{Healthy~(Z)} (20 times).
Therefore, the strength of the transition co-occurrence $((C,D) \Rightarrow_\trans (W,W))$ is $\frac{100}{190}$.

\emph{Forward-looking co-occurrence} $(\state_i \wedge \state_j \Rightarrow_\forw (\state_j,\state'_j))$ is defined as the conditional probability that the next transition executed in $\pers_j$ goes to state $\state'_j$, given that $\pers_j$ is in state $\state_j$ and that $\pers_i$ is in state $\state_i$ during and after the next transition in $\pers_j$.
The strength of this interaction is calculated as the number of times we observe a transition from $\state_j$ to $\state'_j$ while $\pers_i$ has the specified state $\state_i$ divided by the total number of outgoing transitions from $\state_j$ while $\pers_i$ is in $\state_i$.

\begin{definition}\label{def:forwardinteraction}
Let $\sys$ be a \csm~with artifacts $i$ and $j$, $\state_i \in \states_i$, and $\state_j,\state'_j \in \astates_j$, $\state_j \neq \state'_j$.
The strength of the forward-looking co-occurrence $(\state_i \wedge \state_j \Rightarrow_\forw (\state_j,\state'_j))$ is defined as:
\begin{align*}
\hat{P}_\forw(\state_i \wedge \state_j \Rightarrow_\forw & (\state_j,\state'_j)) = \\
    & \frac{\tfreq(((\state_i,\state_j),(\state_i,\state'_j)),\execs_\sys^{\{ i,j \}})}
    {\sum_{\state''_j \in \states_j} \tfreq(((\state_i,\state_j),(\state_i,\state''_j)),\execs_\sys^{\{ i,j \}})}
\end{align*}
\end{definition}

In \autoref{fig:annotatedRunningExample} there are transitions from \emph{Waiting on result~(C)} to \emph{New test needed~(E)} that occur while the patient is \emph{In treatment~(Y)} (10  times).
While \emph{In treatment~(Y)} and \emph{Waiting on result~(C)} there are also transitions to \emph{Result ready~(D)} (70 times).
Therefore the interaction $(Y \wedge C \Rightarrow_\forw (C,E))$ has a strength of $\frac{10}{80}$.

It is possible to calculate the artifact interactions defined above for all pairs of states and transitions of all pairs of artifacts.
However, it is clear that this results in a very large number of interactions for a process analyst to inspect.
One solution to this problem is to rank and filter the list of interactions to obtain the most interesting artifact relations and to present those to the analyst first.

\section{Artifact Interaction Interestingness}\label{sec:metrics}

%\change{Explanation of measures of interestingness: page 5-7.}

%Automated guidance is difficult without a good definition of relevant or unexpected and interesting.

In order to rank and filter artifact interactions based on their interestingness it is necessary to be able to quantify ``interestingness''.
As we discussed in \autoref{sec:related}, work has been performed in the field of association rule learning to develop measures of interestingness to help with the analysis of large sets of association rules~\cite{DBLP:journals/is/TanKS04,DBLP:journals/expert/LiuHCM00}.
We have selected a number of such measures and we discuss their meaning and applicability in the context of artifact interactions that represent process behaviour.

\subsection{Probability Interpretation}

The artifact interactions we defined in \autoref{sec:interaction} are implications over binary stochastic variables representing statements of artifact behaviour.
The implications are of the form $(X \Rightarrow Y)$.
Each statement $X$ or $Y$ is either true or false, with a certain probability that can be estimated from process execution data.
The measures of interestingness objectively score statistical correlations between the variables based on these probabilities.
We discuss the probabilities and their interpretations for each type of artifact interaction.

State co-occurrence $(\state_i \Rightarrow_\states \state_j)$ is an implication between stochastic variables of the form $(X_{\state_i} \Rightarrow Y_{\state_j})$ with $X_{\state_i}$ defined as \emph{$\pers_i$ has state $\state_i$} and $Y_{\state_j}$ defined as \emph{$\pers_j$ has state $\state_j$}.
The probability of $X_{\state_i}$ can be estimated based on the total sojourn time over all states:
\[
\hat{P}_\states(X_{\state_i}) = \frac{\sojourn(\state_i, \execs_\sys^{\{ i \}})}{\sum_{\state \in \states} \sojourn(\state, \execs_\sys)}
\]

Transition co-occurrence $((\state_i,\state'_i) \Rightarrow_\trans (\state_j,\state'_j))$ is either an implication of the form $(X_{(\state_i,\state'_i)} \Rightarrow Y_{\state_j})$ if $\state_j = \state'_j$, with $X_{(\state_i,\state'_i)}$ defined as \emph{$\pers_i$ is in transition from $\state_i$ to $\state'_i$}, or it is an implication $(X_{(\state_i,\state'_i)} \Rightarrow Y_{(\state_j,\state'_j)})$ if $\state_j \neq \state'_j$.
Strictly speaking, the probability of $X_{(\state_i,\state'_i)}$ cannot be expressed because transitions are instantaneous and on a continuous time scale the probability to be in the specific point in time where the transition occurs is infinitesimal, \ie not distinguishable from $0$.
As a result, a number of measures of interestingness would not be defined for transition co-occurrence.
We express the probability based on the total frequency of transitions to avoid this issue:
\[
\hat{P}_\trans(X_{(\state_i,\state'_i)}) = \frac{\tfreq((\state_i,\state'_i),\execs_\sys^{\{ i \}})}{\sum_{(\state,\state') \in \trans} \tfreq((\state,\state'),\execs_\sys)}
\]
%The probability of $Y$ can be estimated if $\state_j = \state'_j$ by $\hat{P}_\states(Y)$, and by $\hat{P}_\trans(Y)$ if $\state_j \neq \state'_j$.

Forward-looking co-occurrence $(\state_i \wedge \state_j \Rightarrow_\forw (\state_j,\state'_j))$ is of the form $(X_{\state_i \wedge \state_j} \Rightarrow Y_{\forw(\state_j,\state'_j)})$ with $X_{\state_i \wedge \state_j}$ defined as \emph{$\pers_j$ has state $\state_j$ and $\pers_i$ has state $\state_i$ during the next transition in $\pers_j$}, and $Y_{\forw(\state_j,\state'_j)}$ defined as \emph{the next transition in $\pers_j$ is from $\state_j$ to $\state'_j$}.
The probability of $X_{\state_i \wedge \state_j}$ is estimated by the probability to be in $\state_j$ and the frequency of $\state_i$ in all possible transitions from $\state_j$:
\begin{align*}
\hat{P}_\forw(X_{\state_i \wedge \state_j}) = &
    \frac{\sojourn(\state_j, \execs_\sys^{\{ j \}})}{\sum_{\state \in \states} \sojourn(\state, \execs_\sys)} * \\
    & \frac{\sum_{\state''_j \in \states_j} \tfreq(((\state_i,\state_j),(\state_i,\state''_j)),\execs_\sys^{\{ i,j \}})}{\sum_{\state''_j \in \states_j} \tfreq((\state_j,\state''_j),\execs_\sys^{\{ j \}})}
\end{align*}
Because $Y_{\forw(\state_j,\state'_j)}$ is only possible if $\pers_j$ has state $\state_j$ we can estimate it with the probability to be in $\state_j$ and the frequency of each possible outgoing transition from $\state_j$:
\begin{align*}
\hat{P}_\forw(Y_{\forw(\state_j,\state'_j)}) = &
    \frac{\sojourn(\state_j, \execs_\sys^{\{ j \}})}{\sum_{\state \in \states} \sojourn(\state, \execs_\sys)} * \\
    & \frac{\tfreq((\state_j,\state'_j),\execs_\sys^{\{ j \}})}{\sum_{\state''_j \in \states_j} \tfreq((\state_j,\state''_j),\execs_\sys^{\{ j \}})}
\end{align*}

\subsection{Measures of Interestingness}

Below we present a selection of measures of interestingness that have been implemented in the \plugin~to evaluate the interestingness of artifact interactions.
The motivation for this selection is that each of these measures has an intuitive interpretation, and that evaluation studies in other application areas have shown that these measures have high predictive power and low collinearity with each other when used to approximate association rule interestingness~\cite{DBLP:conf/edm/BazalduaBP14}.

For each measure we provide a definition, a short description of its intuitive meaning and its range.
The measures are defined in terms of the probabilities of observing the conditions and consequences of the implications representing the different types of artifact interaction.
Some measures are symmetric, \ie their value for $X \Rightarrow Y$ is equal for $Y \Rightarrow X$.

\subsubsection*{Confidence}\label{sec:confidence}

The confidence of an artifact interaction is also referred to as the strength of the prediction, which we introduced for each type of artifact interaction in \autoref{sec:interaction}.
It is defined as a conditional probability:

\begin{equation*}\label{eq:confArt}
  \conf(X \Rightarrow Y) = P(X \Rightarrow Y) = P(Y | X) %= \frac{\supp(X \Rightarrow Y)}{\supp(X)} = \frac{P(X \wedge Y)}{P(X)}
\end{equation*}

The range of $\conf$ is $[0,1]$ and it is asymmetric, \ie in general $\conf(X \Rightarrow Y) \neq \conf(Y \Rightarrow X)$.

\subsubsection*{Support}\label{sec:support}

In the context of association rule learning the support measure is traditionally defined as the frequency with which items occur in a set of transactions, which is an estimate of their probability of occurrence.
In the setting of artifact interaction the support of individual statements is their probability interpretation as defined in the section above, \eg $\supp(X_{\state_i}) = \hat{P}_\states(X_{\state_i})$ and $\supp(Y_{\forw(\state_j,\state'_j)}) = \hat{P}_\forw(Y_{\forw(\state_j,\state'_j)})$.
The support of an implication $X \Rightarrow Y$ is then the probability that the implication is true, multiplied by the probability of observing the condition of the implication:

\begin{equation*}\label{eq:supArt}
  \supp(X \Rightarrow Y) = P(Y | X)P(X) = P(X \wedge Y)
\end{equation*}

The range of $\supp$ is $[0,1]$ and it is symmetric.

\subsubsection*{Lift}\label{sec:lift}

The lift of an interaction is defined as the ratio between the probabilitiy of co-occurrence and the expected co-occurrence under statistical independence:

\begin{equation*}\label{eq:liftArt}
  \lift(X \Rightarrow Y) = \frac{\conf(X \Rightarrow Y)}{\supp(Y)} = \frac{P(Y | X)}{P(Y)} = \frac{P(X \wedge Y)}{P(X)P(Y)}
\end{equation*}

The range of $\lift$ is $[0,\infty]$ and it is symmetric.
A lift measure of $0$ indicates that they are never observed together, a value of $1$ indicates that $X$ and $Y$ are independent, and a value above $1$ indicates that $X$ and $Y$ are observed together more often than can be expected under conditions of statistical independence.

\subsubsection*{Conviction}\label{sec:conviction}

The conviction of an interaction is similar to lift, but it is a directed measure.
It looks at the expected probability of observing $X$ without $Y$, \ie the frequency of the implication being incorrect.
It is defined as the ratio of the frequency of the implication being incorrect, if they were statistically independent, and the frequency of actual observations of the implication not holding:

\begin{equation*}\label{eq:convArt}
  \conv(X \Rightarrow Y) = \frac{1-\supp(Y)}{1-\conf(X \Rightarrow Y)} = \frac{P(X)P(\overline{Y})}{P(X \wedge \overline{Y})}
\end{equation*}

The range of $\conv$ is $(0,\infty]$ and it is asymmetric.
A conviction measure of $1$ indicates that $X$ and $Y$ are statistically independent, while a measure value of $\infty$ occurs for interactions that always hold in the observed data.

\subsubsection*{Cosine}\label{sec:cosine}

The cosine measure is defined as the geometric mean of lift and support:

\begin{equation*}\label{eq:cosArt}
  \cosi(X \Rightarrow Y) = \frac{\supp(X \Rightarrow Y)}{\sqrt{\supp(X)\supp(Y)}} = \frac{P(X \wedge Y)}{\sqrt{P(X)P(Y)}}
\end{equation*}

The range of $\cosi$ is $[0,1]$ and it is symmetric.
It is a null-invariant measurement, \ie it is not affected by the number of observations involving neither $X$ nor $Y$ in the dataset, while \eg the lift measure does not have this property.

\subsubsection*{Jaccard}\label{sec:jaccard}

The jaccard of an interaction is defined as the ratio between the probability of the co-occurrence of $X$ and $Y$ and the probability of observing either:

\begin{align*}\label{eq:jacArt}
  \jacc(X \Rightarrow Y) & = \frac{\supp(X \Rightarrow Y)}{\supp(X) + \supp(Y) - \supp(X \Rightarrow Y)} \\
  & = \frac{P(X \wedge Y)}{P(X \vee Y)}
\end{align*}

The range of $\jacc$ is $[0,1]$, it is symmetric and a null-invariant measurement.
A jaccard measure of $0$ means that items from $X$ and $Y$ are never observed together, and a value of $1$ indicates that if they occur then they are always observed together.

\subsubsection*{Phi-coefficient}\label{sec:phi}

The $\phim$-coefficient of an interaction is defined as the normalised difference between the probability of co-occurrence and the expected probability of co-occurrence under statistical independence:

\begin{equation*}\label{eq:phiArt}
  \phim(X \Rightarrow Y) = \frac{P(X \wedge Y) - P(X)P(Y)}{\sqrt{P(X)P(Y)(1-P(X))(1-P(Y))}}
\end{equation*}

The range of $\phim$ is $[-1,1]$ and it is symmetric.
A value of $0$ indicates that $X$ and $Y$ are statistically independent.

\section{Analysis Guidance Implementation}\label{sec:implementation}

In this section we discuss the implementation of the analysis guidance in the \plugin~\cite{DBLP:conf/bpm/EckSA16a}, a plug-in\footnote{Contained in the \emph{\package}~package of the ProM 6 nightly build, available at {\url{http://www.promtools.org/}}.} in the process mining framework ProM.

The \plugin~discovers a model of the artifact system and of each artifact in the input log, annotates them with sojourn times and frequencies, and presents them in an interactive visualisation.
The interaction allows the user to click on a state or transition and this will highlight all other states and transitions for which $\supp(X \Rightarrow Y) > 0$, based on either $\hat{P_\states}$, $\hat{P_{\trans}}$ or $\hat{P_\forw}$.
The colour of the highlighting is dependent on $\conf(X \Rightarrow Y)$.

\begin{figure*}[bt]
\centering
\includegraphics[width=0.99\textwidth]{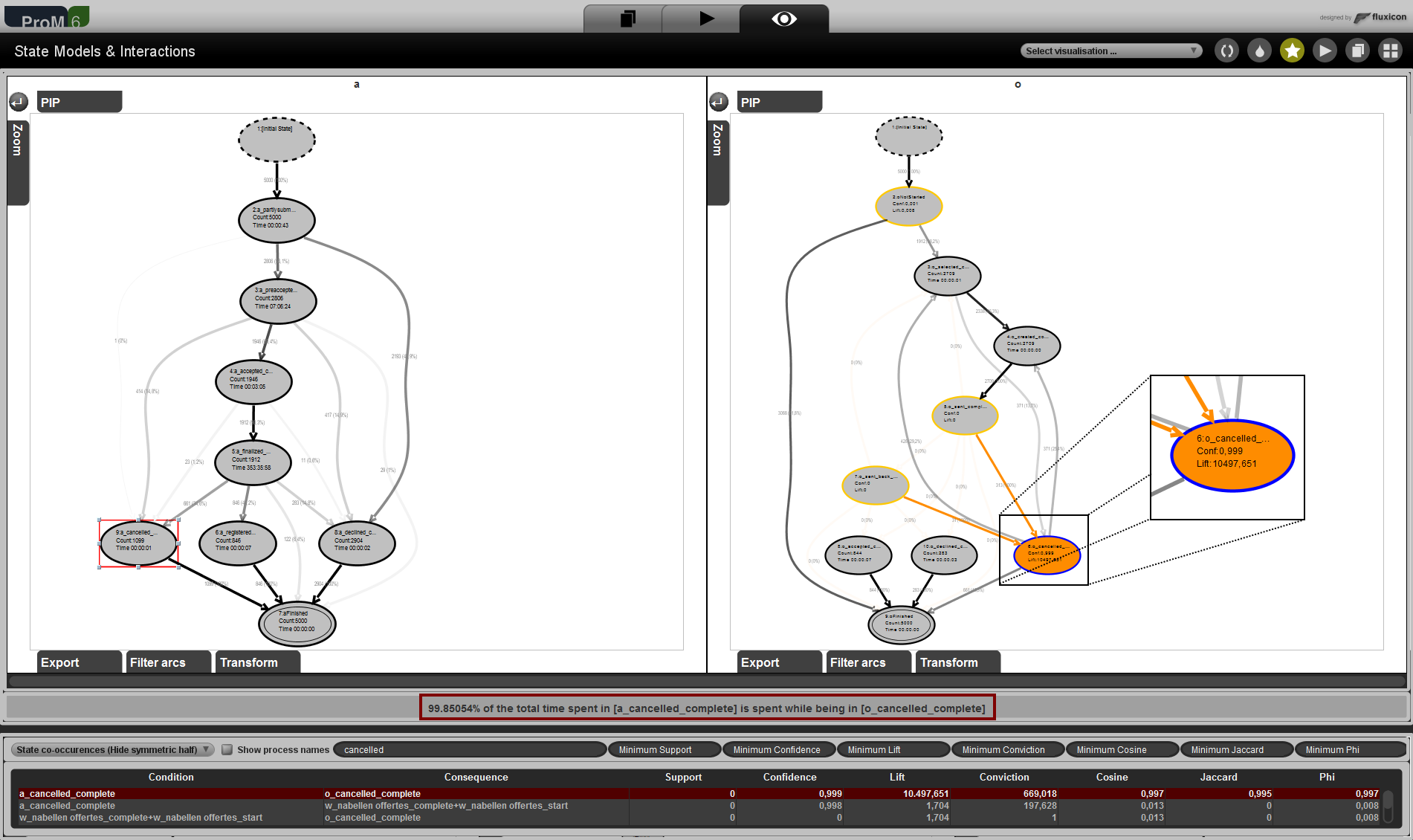}
\caption{The analysis guidance is shown below the process models. Users can sort and filter on the different measures of interestingness, and then click on an artifact interaction to highlight it. The highlighted interaction is also presented as a textual interpretation.}
\label{fig:gui}
\end{figure*}

The analysis guidance for the exploration of artifact interactions is provided below the interactive model visualisation, as shown in \autoref{fig:gui}.
It provides a list of artifact interactions and for each interaction the measures discussed in \autoref{sec:metrics} are calculated.
The user can sort the interactions by the measure values and can set minimum values for each measure to filter the list.

When clicking on the artifact interactions in the list, the user is also presented with a textual interpretation based on four possible templates:

\begin{itemize}
    \item ``$\conf(\state_i \Rightarrow_\states \state_j)$ of the total time spent in $\state_i$ is spent while being in $\state_j$'' (state co-occurrence)
    \item ``Transitions from $\state_i$ to $\state'_i$ occur $\conf((\state_i,\state'_i) \Rightarrow_\trans \state_j)$ of the times while being in $\state_j$'' (transition co-occurrence)
    \item ``Transitions from $\state_i$ to $\state'_i$ occur $\conf((\state_i,\state'_i) \Rightarrow_\trans (\state_j,\state'_j))$ of the times together with a transition from $\state_j$ to $\state'_j$'' (transition co-occurrence)
    \item ``A transition from $\state_j$ goes $\conf(\state_i \wedge \state_j \Rightarrow_\forw (\state_j,\state'_j))$ of the times to $\state'_j$ while being in $\state_i$ (compared to $\hat{P}_\forw((\state_j,\state'_j)|\state_j)$ on average)'' (forward-looking co-occurrence)
\end{itemize}

%\diff{TODO: discuss the GUI, limit association rules to pairs of single states or transitions, give list with 4 templates of interaction, optional automated ranking?}

\section{Evaluation}\label{sec:evaluation}

%\change{This is the evaluation: page 7-8.}

To be able to create analysis guidance that suggests artifact interactions of interest to process analysts it is necessary to establish what qualifies as interesting or relevant.
The measures introduced in \autoref{sec:metrics} are suggestions to quantify interest from the field of association rule learning, but it is unclear how these measures relate to the actual judgments of interest by process analysts.
The aim of this evaluation is to show that the analysis guidance highlights behaviour in real life processes that is useful for understanding the process.
Therefore, we compare the suggestions provided by the tool with insights obtained by other researchers using traditional process mining approaches on real life process data.

\subsection{Process Description}

The event data was taken from the BPI Challenge 2012~\cite{BPIChallenge2012}.
This dataset concerns process instances of a personal loan and overdraft application process at a Dutch financial institute.
The events and activities in the log are related to three interrelated sub-processes, which can be considered as interacting process artifacts.
The first artifact concerns the state of the application (\emph{A}-states), the second relates to the work-items performed by the financial institute (\emph{W}-states), and the third concerns the state of a potential offer that the institute can make to the applicant (\emph{O}-states).
This process has been analysed in several other papers~\cite{DBLP:conf/bpm/BautistaWA12,DBLP:conf/bpm/AdriansyahB12}.

The overall process behavior is as follows.
The process starts with the submission of the application.
An unlogged check determines whether the application is pre-accepted or declined immediately.
The application is accepted once all necessary information has been provided to complete the application.
After the acceptance, the institute sends a concrete offer  for the terms of the loan or overdraft to the applicant.
When the response is returned, the application is validated and then accepted or declined.
At any point in the process the applicant can decide to cancel their application and exit the process.
In cases where the applicant does not respond in a timely manner, or if the application does not meet the criteria of the financial institute, then the application can be declined by the institute.
In exceptional cases the financial institute checks the applications for fraud.

\subsection{Results}

The data of the above process was mined by the \plugin~and then analysed by looking at the measures of interestingness.
We present a list of the top artifact interactions for several of the measures from \autoref{sec:metrics} and explain their relevance for understanding the process behaviour.
Such lists can be obtained in the tool by sorting on the desired measure.

\autoref{tab:topConf} shows five examples of state co-occurrences with high \conf~scores.
There are several state co-occurrences that have a \conf~score of $1$, indicating that a given artifact state always co-occurs with a single state in another artifact.
Not all of these are shown here because most are the result of the offer artifact not changing state from \emph{o::notStarted} until after the application has been accepted.
In general, the state co-occurrences with a high \conf~score indicate relations between artifact states that match the expected flow of the process as also described in other work~\cite{DBLP:conf/bpm/BautistaWA12}.
For example, if the loan is activated then the offer has been accepted by the customer (\emph{a::activated}~$\Rightarrow_\states$~\emph{o::accepted}), and if the application is approved then the application has been validated (\emph{a::approved}~$\Rightarrow_\states$~\emph{w::validation.end}).
Another example is that the financial institute only contacts the customer to follow-up on an offer after the offer has been sent (\emph{w::followupOffers.start}~$\Rightarrow_\states$~\emph{o::sent}).
State co-occurrences highlighted with high \conf~scores can be compared to concurrent dependencies between events or activities in traditional process mining.

\begin{table}
\centering
\caption{Top \conf~State Co-occurrence.}
\label{tab:topConf}

\begin{tabular}
	{ l | l | c }
	Condition & Consequence & $\conf$ \\
    \hline
    a::accepted & o::notStarted & 1 \\
    w::processLeads.start & o::notStarted & 1 \\
    a::activated & o::accepted & 1 \\
    a::approved & w::validation.end & 0.998 \\
    w::followupOffers.start & o::sent & 0.986 \\
\end{tabular}

\end{table}

By contrast, a high \supp~measure indicates the co-occurring artifact states where a lot of time is spent.
\autoref{tab:topSupp} shows the five pairs of artifact states with the highest \supp~scores; note that \supp~is a symmetric measure so condition and consequence are interchangeable.
These results show that almost two thirds of the average time spent in a loan application is spent waiting for the customer to respond after the application has been finalised (\emph{a::finalized}~$\Rightarrow_\states$~\emph{o::sent}).
During this period some time is spent calling the customer, but most of it is spent in between follow-ups (\emph{a::finalized}~$\Rightarrow_\states$~\emph{w::followupOffers.end}).
Additionally, this measure shows that around $20\%$ of the average total time is spent completing the application before an offer is sent out (\emph{a::preaccepted}~$\Rightarrow_\states$~\emph{o::notStarted}).
These imbalances indicate a potential bottleneck at the customer.
This shows that an initial overview of this measure can point out performance issues and encourage a process analyst to do a more thorough process performance and bottleneck analysis.
The insights also match results from other process analyses~\cite{DBLP:conf/bpm/BautistaWA12,DBLP:conf/bpm/AdriansyahB12}.

\begin{table}
\centering
\caption{Top 5 \supp~State Co-occurrence.}
\label{tab:topSupp}

\begin{tabular}
	{ l | l | c }
	Condition & Consequence & $\supp$ \\
    \hline
    a::finalized & o::sent & 0.657 \\
    a::finalized & w::followupOffers.end & 0.577 \\
    w::followupOffers.end & o::sent & 0.504 \\
    a::preaccepted & o::notStarted & 0.191 \\
    w::completeApplication.end & o::notStarted & 0.175
\end{tabular}

\end{table}

A top five of most interesting artifact interactions according to the \lift~measure is shown in \autoref{tab:topLift}.
\lift~shows relations between artifacts that are statistically strong in the sense that they occur much more often than would be expected by chance under independence assumptions.
These results indicate that there are different reasons for an application to be declined (\emph{a::declined}): the offer may have been declined (\emph{o::declined}), the lead may not have matched the required criteria (\emph{a::declined}~$\Rightarrow_\states$~\emph{w::processLeads.start}), or fraud may have been discovered (\emph{a::declined}~$\Rightarrow_\states$~\emph{w::fraudDetection.start}).
It also highlights the synchronisation between the cancellation of the application and the offer (\emph{a::cancelled}~$\Rightarrow_\states$~\emph{o::cancelled}), as \lift~is a symmetric.
Although the results are not surprising given the process description and semantic understanding of the state names, \lift~does provide understanding of the strongest connections between artifacts.

\begin{table}
\centering
\caption{Top 5 \lift~State Co-occurrence.}
\label{tab:topLift}

\begin{tabular}
	{ l | l | c }
	Condition & Consequence & $\lift$ \\
    \hline
    a::declined & o::declined & 596207 \\
    a::cancelled & o::cancelled & 10498 \\
    a::declined & w::processLeads.start & 834 \\
    w::validation.start & o::declined & 819 \\
    a::declined & w::fraudDetection.start & 626 \\
\end{tabular}

\end{table}

Similar to \lift, \conv~also provides an overview of strong relations between artifacts, but this measure is asymmetric in condition and consequence.
\autoref{tab:topConv} shows several relations with high conviction.
We have omitted relations that have even higher \conv~scores but that were also highlighted by the other measures.
Again, the results show relations that are consistent with other analyses~\cite{DBLP:conf/bpm/BautistaWA12}.
For example, given that the application has been validated we know that the offer must have been sent back, and given that the application has been accepted or preaccepted we know that the customer must have provided information to complete the application.

\begin{table}
\centering
\caption{Top \conv~State Co-occurrence.}
\label{tab:topConv}

\begin{tabular}
	{ l | l | c }
	Condition & Consequence & $\conv$ \\
    \hline
    w::validation.end & o\::sentBack & 10.4 \\
    a::accepted & w::completeApplication.end & 10.0 \\
    a::preaccepted & w::completeApplication.end & 7.61 \\
    a::activated & w::validation.end & 7.54 \\
    w::callIncompleteFiles.start & o::sentBack & 6.70 \\
\end{tabular}

\end{table}

The top results for state co-occurrence in terms of $\cosi$, $\jacc$ and $\phim$ generally score high on at least one other measure.
The exact order of the artifact interactions differs between the measures, but in general the state co-occurrence relations that are scored as most interesting are those that have a strong link to the overall behaviour of the application process.
%\diff{Give an example for each measure for which this is not true, but state that they are not top scores?}

%Top 5 \cosi~state co-occurrence (deduplicated with previous).
%
%\begin{table}
%\centering
%\caption{Top 5 \cosi~state co-occurrence.}
%\label{tab:topCosi}
%
%\begin{tabular}
%	{ c | c | c }
%	Condition & Consequence & $\cosi$ \\
%    \hline
%    a\::finalized & o\::sent & 1 \\
%\end{tabular}
%
%\end{table}
%
%Top 5 \jacc~state co-occurence.
%
%\begin{table}
%\centering
%\caption{Top 5 \jacc~state co-occurrence.}
%\label{tab:topJacc}
%
%\begin{tabular}
%	{ c | c | c }
%	Condition & Consequence & $\jacc$ \\
%    \hline
%    a\::finalized & o\::sent & 1 \\
%\end{tabular}
%
%\end{table}
%
%Top 5 $\phim$~state co-occurrence.
%
%\begin{table}
%\centering
%\caption{Top 5 $\phim$~state co-occurrence.}
%\label{tab:topPhi}
%
%\begin{tabular}
%	{ c | c | c }
%	Condition & Consequence & $\phim$ \\
%    \hline
%    a\::finalized & o\::sent & 1 \\
%\end{tabular}
%
%\end{table}

\autoref{tab:topTConf} shows several transitions that always co-occur with the application state \emph{a::finalized}.
This means that these transitions, such as the creation and sending of an offer, are only enabled if the application has been finalized ((\emph{o::created,o::sent})~$\Rightarrow_\trans$~\emph{a::finalized}), \ie if all the required information has been provided.
In general, there are many trivial transition co-occurrences that have a \conf~of $1$, which means there are clear synchronization points in the interaction between the artifacts.
Other examples are related to the start of the process that only involves the application artifact.

\begin{table}
\centering
\caption{Top \conf~Transition Co-occurrence.}
\label{tab:topTConf}

\begin{tabular}
	{ l | l | c }
	Condition & Consequence & $\conf$ \\
    \hline
    from o::selected, to o::cancelled & a::finalized & 1 \\
    from o::selected, to o::created & a::finalized & 1 \\
    from o::created, to o::sent & a::finalized & 1 \\
    from w::completeApplication.end, & \multirow{2}{*}{a::finalized} & \multirow{2}{*}{1} \\
        \quad to w::followupOffers.start & & \\
    from w::fraudDetection.end, & \multirow{2}{*}{a::finalized} & \multirow{2}{*}{1} \\
        \quad to w::validation.start & & \\
\end{tabular}

\end{table}

There are many transition co-occurrences with high \lift~metric scores due to the clear synchronisation between artifacts.
\autoref{tab:topTLift} shows a number of these, with a minimum support of $0.001$ to filter out patterns that are the result of very rare transitions.
Especially the strong links between the outcome of the application and the state of the offer are very clear again.
Interestingly, there are transitions from the sending of the offer directly to its acceptance, without receiving a reply to the offer (\emph{o::sentBack}).
The \lift~measure shows that these transitions co-occur significantly often while calling the customer for incomplete information ((\emph{o::sent,o::accepted})~$\Rightarrow_\trans$~\emph{w::callIncompleteFiles.start}).
This shows that it appears that the institute also allows the offer to be verbally accepted by customers during contact by phone.
Also, a significant number of offers that were sent back and then cancelled were cancelled during contact by phone ((\emph{o::sentBack,o::cancelled})~$\Rightarrow_\trans$~\emph{w::callIncompleteFiles.start}).
These observations are not immediately clear when looking at the control flow using traditional approaches~\cite{DBLP:conf/bpm/BautistaWA12,DBLP:conf/bpm/AdriansyahB12}

\begin{table}
\centering
\caption{Top \lift~Transition Co-occurrence.}
\label{tab:topTLift}

\begin{tabular}
	{ l | l | c }
	Condition & Consequence & $\lift$ \\
    \hline
    from o::sent, to o::declined & a::declined & 130 \\
    from o::sent, to o::accepted & a::approved & 54.0 \\
    from o::sent, to o::accepted & w::callIncompleteFiles.start & 18.3 \\
    from w::followupOffers.end, & \multirow{2}{*}{o::sentBack} & \multirow{2}{*}{6.88} \\
        \quad to w::validation.start & & \\
    from o::sentBack, to o::cancelled & w::callIncompleteFiles.start & 4.90 \\
\end{tabular}

\end{table}

The above discussion shows that the presented approach is able to highlight artifact interactions that provide insights into the behaviour of a real life process.
The insights obtained are comparable with those provided by traditional process mining approaches, but they do not require an analysis of the control flow of a complex or unstructured process model.
Sorting and filtering functionalities ensure that the size of the list of potentially interesting artifact interactions remains manageable.
However, there are often interactions that score well on multiple measures and it currently remains up to the user to identify the overlap between the top scoring interactions for two or more measures.

\section{Conclusion \& Future Work}\label{sec:conclusion}

%\change{This is the conclusion: page 11.}

In this paper we have presented an approach to objectively quantify the interestingness of interactions between artifacts in artifact-centric processes.
This approach is based on measures of interestingness that have been defined in the context of process models.
It highlights useful or surprising artifact interactions and thereby enables process analysts to deal with large or complex models.
The approach has been implemented using an interactive process discovery tool, the \plugin, which has been shown to provide relevant insights on real life process execution data.
Most of the insights discussed can also be obtained with traditional process mining techniques, but they require data preprocessing to obtain structured models and careful analysis of the behaviour of those complex models.

We aim to extend this work in several ways.
The current evaluation is limited and provides only an indication of the usefulness of the approach in practice.
We plan to conduct a user study to relate the objective measures of interestingness to the subjective interests of process analysts.
Controlled experiments could also provide indications for cut-off or minimal values for the measures.

Extensions of the approach itself are also possible.
Instead of only looking at pairs of artifacts, we can generalise artifact interaction to sets of artifacts, similar to association rule learning.
In contrast to association rule learning, infrequent relations may also be interesting when analysing a process.
There is also room to improve the transformation of execution sequences into observations of artifact interaction.
For example, correlations based on time intervals could be used to handle noise or non-fitting executions in the process data.

%\change{References: page 11-12.}

%\bibliographystyle{plain}
%\bibliographystyle{splncs03}
%\bibliography{lit}
\bibliographystyle{IEEEtran}
\bibliography{IEEEabrv,lit}

\begin{thebibliography}{10}
\providecommand{\url}[1]{#1}
\csname url@samestyle\endcsname
\providecommand{\newblock}{\relax}
\providecommand{\bibinfo}[2]{#2}
\providecommand{\BIBentrySTDinterwordspacing}{\spaceskip=0pt\relax}
\providecommand{\BIBentryALTinterwordstretchfactor}{4}
\providecommand{\BIBentryALTinterwordspacing}{\spaceskip=\fontdimen2\font plus
\BIBentryALTinterwordstretchfactor\fontdimen3\font minus
  \fontdimen4\font\relax}
\providecommand{\BIBforeignlanguage}[2]{{%
\expandafter\ifx\csname l@#1\endcsname\relax
\typeout{** WARNING: IEEEtran.bst: No hyphenation pattern has been}%
\typeout{** loaded for the language `#1'. Using the pattern for}%
\typeout{** the default language instead.}%
\else
\language=\csname l@#1\endcsname
\fi
#2}}
\providecommand{\BIBdecl}{\relax}
\BIBdecl

\bibitem{Aalst2016Book}
W.~M.~P. van~der Aalst, \emph{Process Mining - Data Science in Action, Second
  Edition}.\hskip 1em plus 0.5em minus 0.4em\relax Springer, 2016.

\bibitem{DBLP:conf/bpm/EckSA16}
M.~L. van Eck, N.~Sidorova, and W.~M.~P. van~der Aalst, ``Discovering and
  exploring state-based models for multi-perspective processes,'' in
  \emph{Business Process Management - 14th International Conference, {BPM}
  2016, Rio de Janeiro, Brazil, September 18-22, 2016. Proceedings}, 2016, pp.
  142--157.

\bibitem{DBLP:journals/ijcis/AalstBEW01}
W.~M.~P. van~der Aalst, P.~Barthelmess, C.~A. Ellis, and J.~Wainer, ``Proclets:
  {A} framework for lightweight interacting workflow processes,'' \emph{Int. J.
  Cooperative Inf. Syst.}, vol.~10, no.~4, pp. 443--481, 2001.

\bibitem{DBLP:journals/ijcis/PopovaFD15}
V.~Popova, D.~Fahland, and M.~Dumas, ``Artifact lifecycle discovery,''
  \emph{Int. J. Cooperative Inf. Syst.}, vol.~24, no.~1, 2015.

\bibitem{DBLP:journals/tsc/LuNWF15}
X.~Lu, M.~Nagelkerke, D.~van~de Wiel, and D.~Fahland, ``Discovering interacting
  artifacts from {ERP} systems,'' \emph{{IEEE} Trans. Services Computing},
  vol.~8, no.~6, pp. 861--873, 2015.

\bibitem{DBLP:conf/bpm/PopovaD13}
V.~Popova and M.~Dumas, ``Discovering unbounded synchronization conditions in
  artifact-centric process models,'' in \emph{Business Process Management
  Workshops - {BPM} 2013 International Workshops, Beijing, China, August 26,
  2013, Revised Papers}, 2013, pp. 28--40.

\bibitem{DBLP:journals/is/TanKS04}
P.~Tan, V.~Kumar, and J.~Srivastava, ``Selecting the right objective measure
  for association analysis,'' \emph{Inf. Syst.}, vol.~29, no.~4, pp. 293--313,
  2004.

\bibitem{DBLP:journals/expert/LiuHCM00}
B.~Liu, W.~Hsu, S.~Chen, and Y.~Ma, ``Analyzing the subjective interestingness
  of association rules,'' \emph{{IEEE} Intelligent Systems}, vol.~15, no.~5,
  pp. 47--55, 2000.

\bibitem{DBLP:conf/bpm/EckSA16a}
M.~L. van Eck, N.~Sidorova, and W.~M.~P. van~der Aalst, ``Composite state
  machine miner: Discovering and exploring multi-perspective processes,'' in
  \emph{Proceedings of the {BPM} Demo Track 2016 Co-located with the 14th
  International Conference on Business Process Management {(BPM} 2016), Rio de
  Janeiro, Brazil, September 21, 2016.}, 2016, pp. 73--77.

\bibitem{DBLP:conf/caise/BoseVA11a}
R.~P. J.~C. Bose, H.~M. W.~E. Verbeek, and W.~M.~P. van~der Aalst,
  ``Discovering hierarchical process models using prom,'' in \emph{{IS}
  Olympics: Information Systems in a Diverse World - CAiSE Forum 2011, London,
  UK, June 20-24, 2011, Selected Extended Papers}, 2011, pp. 33--48.

\bibitem{DBLP:journals/tkde/WeerdtBVB13}
J.~D. Weerdt, S.~K. L.~M. vanden Broucke, J.~Vanthienen, and B.~Baesens,
  ``Active trace clustering for improved process discovery,'' \emph{{IEEE}
  Trans. Knowl. Data Eng.}, vol.~25, no.~12, pp. 2708--2720, 2013.

\bibitem{DBLP:conf/edm/BazalduaBP14}
D.~L. Bazaldua, R.~S. Baker, and M.~O.~S. Pedro, ``Comparing expert and
  metric-based assessments of association rule interestingness,'' in
  \emph{Proceedings of the 7th International Conference on Educational Data
  Mining, {EDM} 2014, London, UK, July 4-7, 2014}, 2014, pp. 44--51.

\bibitem{BPIChallenge2012}
\BIBentryALTinterwordspacing
B.~F. van Dongen, ``Bpi challenge 2012,'' 2012. [Online]. Available:
  \url{http://dx.doi.org/10.4121/uuid:3926db30-f712-4394-aebc-75976070e91f}
\BIBentrySTDinterwordspacing

\bibitem{DBLP:conf/bpm/BautistaWA12}
A.~D. Bautista, L.~Wangikar, and S.~M.~K. Akbar, ``Process mining-driven
  optimization of a consumer loan approvals process - the {BPIC} 2012 challenge
  case study,'' in \emph{Business Process Management Workshops - {BPM} 2012
  International Workshops, Tallinn, Estonia, September 3, 2012. Revised
  Papers}, 2012, pp. 219--220.

\bibitem{DBLP:conf/bpm/AdriansyahB12}
A.~Adriansyah and J.~C. A.~M. Buijs, ``Mining process performance from event
  logs,'' in \emph{Business Process Management Workshops - {BPM} 2012
  International Workshops, Tallinn, Estonia, September 3, 2012. Revised
  Papers}, 2012, pp. 217--218.

\end{thebibliography}

\end{document}